\documentclass{article}
\usepackage{spconf,amsmath,amsfonts,bm,amssymb,epsfig,psfrag,ifthen,color,subfigure}
\usepackage{algorithm,algpseudocode}

\newcommand\wb{\ensuremath{{\bm w}}}

\newcommand\hb{\ensuremath{{\bf h}}}

\newcommand\Qb{\ensuremath{{\bf Q}}}
\newcommand\qb{\ensuremath{{\bm q}}}

\newcommand\Wb{\ensuremath{{\bf W}}}

\newcommand\zerob{\ensuremath{{\bf 0}}}

\newcommand\tr{\ensuremath{{\rm Tr}}}

\newcommand\rank{\ensuremath{{\rm rank}}}

\newcommand\Cplx{\ensuremath{{\mathbb{C}}}}

\newtheorem{Definition}{Definition}

\ninept
\title{A convex approximation approach to Weighted Sum Rate Maximization of
Multiuser MISO Interference Channel under outage constraints}% Single address.
\name{Wei-Chiang Li, Tsung-Hui Chang, Che Lin, and Chong-Yung Chi
\thanks{This work is supported by National Science Council, R.O.C.,
under Grants NSC 98-2219-E-007-003, NSC 98-2219-E-007-005, NSC
99-2221-E-007-052-MY3 and NSC 98WFA0400537.}}

\address{Institute of Communications Engineering \& Department of Electrical Engineering\\
National Tsing Hua University, \\ Hsinchu, Taiwan 30013 \\
\small E-mail: \{weichiangli, tsunghui.chang\}@gmail.com,~\{clin,
cychi\}@ee.nthu.edu.tw}

\begin{document}
\bibliographystyle{IEEEtran}
\ninept
\maketitle
\begin{abstract}
This paper considers weighted sum rate maximization of multiuser
multiple-input single-output interference channel (MISO-IFC) under
outage constraints. The outage-constrained weighted sum rate
maximization problem is a nonconvex optimization problem and is
difficult to solve. While it is possible to optimally deal with this
problem in an exhaustive search manner by finding all the
Pareto-optimal rate tuples in the (discretized) outage-constrained
achievable rate region, this approach, however, suffers from a
prohibitive computational complexity and is feasible only when the
number of transmitter-receive pairs is small. In this paper, we
propose a convex optimization based approximation method for
efficiently handling the outage-constrained weighted sum rate
maximization problem. The proposed approximation method consists of
solving a sequence of convex optimization problems, and thus can be
efficiently implemented by interior-point methods. Simulation
results show that the proposed method can yield near-optimal
solutions.
\end{abstract}
\begin{keywords}
Multiuser interference channel, weighted sum rate maximization,
outage probability, convex optimization
\end{keywords}
\vspace{-0.20cm}
\section{Introduction}\vspace{-0.2cm}\label{sec:intro}

Recently, interference management for improving spectral
efficiency of wireless multiuser systems has been a research topic
drawing significant attention \cite{Cadambe08}. This paper
considers the $K$-user multiple-input single-output interference
channel (MISO-IFC) where $K$ multi-antenna transmitters
simultaneously communicate with $K$ respective single-antenna
receivers over a common frequency band. This MISO-IFC arises, for
example, in multicell wireless systems where each of the base
stations is equipped with multiple antennas and each mobile
station has only one antenna. Under the assumption that the
transmitters have the perfect channel state information, and that
the receivers employ single-user detection, it has been shown that
transmit beamforming is an optimal transmission scheme to attain
the Pareto boundary of the achievable rate region of MISO-IFC
\cite{Shang_2009}. The structure of the Pareto-optimal beamforming
schemes has also been studied in
\cite{Jorswieck08,Zhang_Cui_2010}. A game-theoretic approach for
MISO-IFC has been presented in \cite{Larsson_etal2009_mag}.

This paper assumes that the channel coefficients are block-faded,
and that the transmitters know only the statistical distribution
of the channels. Specifically, each channel is assumed to be
circularly symmetric complex Gaussian distributed, with a
covariance matrix known to the transmitters. Under limited delay
constraints and due to channel fading, the receivers' performance
may suffer from outage. Assuming that the transmitters employ
transmit beamforming, the achievable rate region of MISO-IFC under
outage constraints on receivers' performance has been investigated
in \cite{Lindblom09}. While this outage-constrained achievable
rate region is not known analytically so far, it has been shown
that this region can be found numerically using an exhaustive
search method \cite{Lindblom09}. This method, unfortunately, has a
complexity that increases exponentially with $K(K-1)$, and
therefore is not feasible in practice.

In this paper, we investigate efficient approaches to achieving
Pareto-optimal beamforming solutions that maximize the achievable
weighted sum rate. To this end, we study the design formulation that
maximizes the weighted sum rate subject to outage constraints and
individual power constraints. Due to the nonconvextity of the outage
constraints, solving the weighted sum rate maximization problem is a
challenging task. To efficiently deal with this problem, we propose
a sequential convex approximation method. The proposed approximation
method is conservative in the sense that the obtained approximate
beamforming solutions are guaranteed to be feasible and satisfy the
outage constraints of the original problem. Since the proposed
method only involves solving convex optimization problems, it can be
efficiently implemented by interior-point methods in a
polynomial-time complexity \cite{cvx}. The presented simulation
results show that the proposed approximation method can provide
near-optimal performance and outperform the existing maximum-ratio
and zero-forcing transmission strategies.

\vspace{-0.2cm}
%%%%%%%%%%%%%%%%%%%%%%%%%%%%%%%%%%%%%%%%%%%%%%%%%%%%%%%%%%%%%%%%%%%%%%%%
\section{Signal Model and Problem Statement}\label{sec:sysmodle}
%%%%%%%%%%%%%%%%%%%%%%%%%%%%%%%%%%%%%%%%%%%%%%%%%%%%%%%%%%%%%%%%%%%%%%%%
\vspace{-0.2cm}

We consider the $K$-user MISO interference channel where each of the
transmitters has $N_t$ antennas and all the receivers are equipped
with a single antenna. All the transmitters employ transmit
beamforming to transmit information signals to their respective
receivers. Let $s_i(t)$ denote the information signal sent from
transmitter $i$, and let $\wb_i \in \Cplx^{N_t}$ be the associated
beamforming vector. The received signal at receiver $i$ is given by
\begin{align}\label{received signal}
r_i(t) = \hb_{ii}^H\wb_is_i(t)+\sum_{k=1,k\neq{i}}^K\hb_{ki}^H\wb_k
s_k(t)+n_i(t),
\end{align}
where $\hb_{ki}\in\mathbb{C}^{N_t}$ denotes the channel vector from
transmitter $k$ to receiver $i$, and $n_i(t)$ is the additive noise
of receiver $i$. The noise $n_i(t)$ is assumed to be complex
Gaussian distributed with zero mean and variance $\sigma^2_i>0$,
i.e., $n_i(t)\sim\mathcal{CN}(0,\sigma_i^2)$. Assuming that
$s_i(t)\sim\mathcal{CN}(0,1)$ and that the receivers decode the
information message using single-user detection (which treats the
cross-link interference as noise), the achievable rate of the $i$th
transmitter-receiver pair is given by
\begin{align*}
r_i\!\left( \{\hb_{ki}\}_{k=1}^K,\{\wb_k\}_{k=1}^K\right)
=\log_2\left(1+\frac{\left|\hb_{ii}^H\wb_i\right|^2}{\sum_{k\neq{i}}\left|\hb_{ki}^H\wb_k\right|^2+\sigma^2_i}\right).
\end{align*}

In this paper, we assume that the channel coefficients $\hb_{ki}$
are block-faded, and that the transmitters can only acquire the
statistical distribution of the channels. In particular, the
elements of $\hb_{ki}$ are assumed to be circularly symmetric
complex Gaussian distributed with covariance matrix equal to
$\Qb_{ki} \succeq \zerob$ (positive semidefinite), i.e.,
$\hb_{ki}\sim\mathcal{CN}(\mathbf{0},\Qb_{ki})$, for all
$k,i=1,\ldots,K$. Let $R_i>0$ be the target transmission rate of
receiver $i$. Due to channel fading, the receivers' performance
may suffer from outage; that is, it would have a nonzero
probability such that
$r_i\!\left(\{\hb_{ki}\}_{k=1}^K,\{\wb_k\}_{k=1}^K\right)<R_i$.
The $\epsilon_i$-outage achievable rate region is defined as
follows:

\begin{Definition} {\rm \cite{Lindblom09}}
Let $P_i>0$ denote the power constraint of transmitter $i$, and let
$\epsilon_i \in (0,1]$ denote the maximum tolerable outage
probability of receiver $i$, for $i=1,\ldots,K$. The rate tuple
$(R_1,\dots,R_K)$ is said to be achievable if
\begin{align*}
\Pr\left\{r_i\!\left(\{\hb_{ki}\}_{k=1}^K,\{\wb_k\}_{k=1}^K\right)<
R_i\right\}\leq \epsilon_i,~i=1,\dots,K
\end{align*}
for some $(\wb_1,\dots,\wb_K) \in \mathcal{W}_1 \times \cdots \times
\mathcal{W}_K$ where
$\mathcal{W}_i\triangleq\{\wb\in\mathbb{C}^{N_t}|~
\|\wb\|^2\le{P_i}\}$. The $\epsilon_i$-outage achievable rate region
is given by
\begin{align*}
&\mathcal{R}=\notag \\
&\bigcup_{\substack{\wb_i\in\mathcal{W}_i, \\
i=1,\dots,K}} \!\!\!\bigg\{ \!\!\!\begin{array}{ll}
(R_1,\dots,R_K)|\!\!\!
&\Pr\left\{r_i\!\left(\{\hb_{ki}\}_{k=1}^K,\{\wb_k\}_{k=1}^K\right)<
R_i\right\}\notag \vspace{0.2cm}\\
&\leq \epsilon_i,~i=1,\dots,K \end{array}\!\!\!\bigg\}.
\end{align*}
\end{Definition}

Given the outage specifications $\epsilon_1,\ldots,\epsilon_K$, it
is desirable to optimize the beamforming vectors $\{\wb_k\}_{k=1}^K$
such that the system can operate on the so-called Pareto boundary of
the achievable rate region $\mathcal{R}$ \cite{Lindblom09}, with
system utilities such as the (weighted) sum of $R_1,\dots,R_K$ being
maximized. To this end, we consider the following weighted sum rate
maximization problem
\begin{center}
\fbox{\parbox[]{0.97 \linewidth}{\vspace{-0.3cm}
\begin{subequations}
\begin{align}
\max_{\substack{\wb_i \in \Cplx^{N_t},R_i\geq 0, \\ i=1,\ldots,K}}~&\sum_{i=1}^K\alpha_iR_i\label{WSR_a}\\
\text{s.t.}~&\Pr\left\{r_i\!\left(\{\hb_{ki}\}_{k=1}^K,\{\wb_k\}_{k=1}^K\right)< R_i\right\}\leq \epsilon_i,\nonumber\\
&~i=1,\dots,K,\label{WSR_b}\\
&~\|\wb_i\|^2\leq P_i,~~i=1,\dots,K,\label{WSR_c}
\end{align}\label{WSR}
\end{subequations}
\vspace{-0.6cm}}}
\end{center}
where $\alpha_i \geq 0$ is the priority weight for the $i$th
transmitter-receiver pair. Solving problem \eqref{WSR} is
challenging because the outage constraints in \eqref{WSR_b} are
difficult to handle. One possible approach to solving problem
\eqref{WSR} is to first obtain a set of Pareto-optimal rate tuples
$(R_1,\dots,R_K)$ by discretizing $\mathcal{R}$ using an
exhaustive search method reported in \cite{Lindblom09}, followed
by picking the one that corresponds to the largest value of
$\sum_{i=1}^K\alpha_iR_i$. The complexity of this approach,
however, increases exponentially with $K(K-1)$\footnote{The
exhaustive search method in \cite{Lindblom09} samples the
achievable rate region $\mathcal{R}$ by discretizing the
cross-link interference into a finite number of levels. Let $M$ be
the number of discretization levels. This method then needs to
list a total number of $M^{K(K-1)}$ rate tuples, and finds the one
with maximum $\sum_{i=1}^K\alpha_iR_i$. For a rough case of $M=10$
and $K=3$, this method requires to search over $10^{6}$ rate
tuples, which is computationally prohibitive.}. In the next
section, based on convex approximation techniques, we present a
suboptimal approach for efficiently handling problem \eqref{WSR}.

%===========================================================================================
\section{Proposed Convex Approximation Method}\label{sec:PropAlgrthm}

\subsection{Closed-Form Expression of Outage
Probability}\label{sec:ClsForm}

While the probability constraints in \eqref{WSR_b} seem
intractable, there actually exist closed-form expressions. To show
this, it is noted that each of the probability in \eqref{WSR_b}
can be expressed as
\begin{align}\label{satisfaction prob}
 \Pr\left\{\frac{\left|\hb_{ii}^H\wb_i\right|^2}{\sum_{k\neq{i}}\left|\hb_{ki}^H\wb_k\right|^2+\sigma^2_i}< 2^{R_i}-1\right\}
\end{align} which is the left tail probability of the ratio of the
exponential random variable $|\hb_{ii}^H\wb_i|^2$ to the sum of
independent exponential random variables $|\hb_{ki}^H\wb_k|^2$ for
$k\neq i$. According to \cite[Appendix I]{Kandukuri02},
\eqref{satisfaction prob} has a closed-form expression as
\begin{align}\label{satisfaction prob2}
\!\!\!1-e^{\frac{-(2^{R_i}-1)\sigma_i^2}{\wb_i^H\Qb_{ii}\wb_i}}
\prod_{k\neq{i}}\frac{\wb_i^H\Qb_{ii}\wb_i}{\wb_i^H\Qb_{ii}\wb_i+(2^{R_i}-1)\wb_k^H\Qb_{ki}\wb_k}.
\end{align}
Hence problem \eqref{WSR} can be equivalently represented by
\begin{align}\label{WSR_CLSFORM}
\max_{\substack{\wb_i \in \Cplx^{N_t},R_i\geq 0, \\ i=1,\ldots,K}}~&\sum_{i=1}^K\alpha_iR_i\\
\text{s.t.}~&\rho_i
e^{\frac{(2^{R_i}-1)\sigma_i^2}{\wb_i^H\Qb_{ii}\wb_i}}\prod_{k\ne{i}}
\left(\!1\!+\!\!\frac{(2^{R_i}-1)\wb_k^H\Qb_{ki}\wb_k}{\wb_i^H\Qb_{ii}\wb_i}\right)\le1,\nonumber\\
&\|\wb_i\|^2\leq P_i,~i=1,\dots,K, \notag
\end{align}
where $\rho_i\triangleq 1-\epsilon_i$. It can be seen that
\eqref{WSR_CLSFORM} is a nonconvex optimization problem. Next, we
show how to approximate problem \eqref{WSR_CLSFORM} by a convex
optimization problem.

\vspace{-0.0cm}
%===========================================================================================
\subsection{Proposed Conservative Formulation}\label{sec:ConsrvApprx}
%===========================================================================================
\vspace{-0.0cm}

The approximation method to be presented is conservative, in the
sense that the obtained approximate solution is guaranteed to be
feasible to problem \eqref{WSR}. To illustrate the proposed
method, let us define
\begin{subequations}\label{change of variables}
\begin{align}
  e^{x_{ki}}&\triangleq \tr(\Wb_k\Qb_{ki}),~~e^{y_{i}}\triangleq 2^{R_i}-1, \label{change of variables a}\\
  z_i&\triangleq
  \frac{2^{R_i}-1}{\tr(\Wb_i\Qb_{ii})}=e^{y_i-x_{ii}}, \label{change of variables b}\\
  \Wb_i
  &\triangleq \wb_i\wb_i^H, \label{change of variables c}
\end{align}
\end{subequations}
where $x_{ki}, y_{i}, z_i \in \mathbb{R}$ are introduced slack
variables for $k,i=1,\ldots,K$, and $\tr(\cdot)$ denotes the trace
of a matrix. Substituting \eqref{change of variables} into
\eqref{WSR_CLSFORM} yields the following problem
\begin{subequations}\label{WSR_ChVar}
\begin{align}
\max_{\substack{\Wb_i \in \mathbb{H}^{N_t},R_i\geq 0, \\ x_{ki},y_i,z_i \in \mathbb{R}, \\k,i=1,\ldots,K}}~&\sum_{i=1}^K\alpha_iR_i,\label{WSR_ChVar_a}\\
\text{s.t.}~&~\rho_ie^{\sigma_i^2z_i}\prod_{k\ne{i}}\left(1+e^{-x_{ii}+x_{ki}+y_i}\right)\leq 1,\label{WSR_ChVar_b}\\
&~\tr(\Wb_i\Qb_{ki})\leq e^{x_{ki}},~~k\in \mathcal{K}^c_i,\label{WSR_ChVar_c}\\
&~\tr(\Wb_i\Qb_{ii})\geq e^{x_{ii}},\label{WSR_ChVar_c2}\\
&~2^{R_i}\leq e^{y_i}+1,\label{WSR_ChVar_d}\\
&~e^{y_i-x_{ii}}\leq z_i,\label{WSR_ChVar_e}\\
&~\tr(\Wb_i)\leq
P_i,\label{WSR_ChVar_f}\\
&~\Wb_i\succeq
\zerob,~\rank(\Wb_i)=1,~i=1,\dots,K,\label{WSR_ChVar_g}
\end{align}
\end{subequations} where $\mathcal{K}^c_i\triangleq
\{1,\ldots,K\}\backslash\{i\}$, and \eqref{WSR_ChVar_g} is due to
\eqref{change of variables c}. Notice that we have replaced the
equalities in \eqref{change of variables a} and \eqref{change of
variables b} with inequalities as in \eqref{WSR_ChVar_c} to
\eqref{WSR_ChVar_e}. It is not difficult to verify that all the
inequalities in \eqref{WSR_ChVar_c} to \eqref{WSR_ChVar_e} would
hold with equalities at the optimum; otherwise a larger optimal
weighted
sum rate can always be obtained. %
%. For example, if \eqref{WSR_ChVar_c2} is inactive at the optimum,
%one can increase ${x_{ii}}$ by a small $\nu>0$ such that
%\eqref{WSR_ChVar_c2} holds with equality for ${x_{ii}}+\nu$. By
%doing so, one can also increase $y_i$ by $\nu>0$ without violating
%\eqref{WSR_ChVar_b} and \eqref{WSR_ChVar_f}. Then by
%\eqref{WSR_ChVar_e}, a higher value of $R_i$ can be obtained. Hence
%\eqref{WSR_ChVar_c2} holds with equality
%
% and altering the
%objective value. This is also true for the inequalities in
%\eqref{WSR_ChVar_c}, \eqref{WSR_ChVar_d} and \eqref{WSR_ChVar_e}.
Therefore, problem \eqref{WSR_ChVar} is equivalent to problem
\eqref{WSR_CLSFORM}.

One can see that the objective function and most of the constraints
of problem \eqref{WSR_ChVar} are convex, except the constraints in
\eqref{WSR_ChVar_c} and \eqref{WSR_ChVar_d}, and the nonconvex
rank-one constraints in \eqref{WSR_ChVar_g}. Let
$(\{\bar{\wb}_i\}_{i=1}^K,\{\bar{R}_i\}_{i=1}^K)$ be a feasible
point of problem \eqref{WSR}. Define
\begin{subequations}\label{feasible point}
\begin{align}
   \bar{x}_{ki}&\triangleq \ln(\bar{\wb}_k^H\Qb_{ki}\bar{\wb}_k),~k\in
   \mathcal{K}^c_{i}, \\
   \bar{y}_i&\triangleq \ln(2^{\bar{R}_i}-1),
\end{align}
\end{subequations} for $i=1,\ldots,K$. Then $\{\{\bar{x}_{ki}\}_{k\neq
i},\bar{y}_i\}_{i=1}^K$ together with $\bar{R}_i$, $\bar{x}_{ii}
\triangleq \ln(\bar{\wb}_i^H\Qb_{ii}\bar{\wb}_i)$, $\bar{\Wb}_i
\triangleq \bar{\wb}_i\bar{\wb}_i^H$ and $\bar{z}_i \triangleq
e^{\bar{y}_i -\bar{x}_{ii}}$ for $i=1,\ldots,K$, are feasible to
problem \eqref{WSR_ChVar}. We aim to conservatively approximate
\eqref{WSR_ChVar_c} and \eqref{WSR_ChVar_d} with respective to the
point $\{\{\bar{x}_{ki}\}_{k\neq i},\bar{y}_i\}_{i=1}^K$. Since
$e^{x_{ki}}$ is convex, its first-order approximation at
$\bar{x}_{ki}$, i.e., ${e}^{\bar{x}_{ki}}(x_{ki}-\bar{x}_{ki}+1)$,
is a global underestimate of $e^{x_{ki}}$. Hence it is sufficient
to achieve \eqref{WSR_ChVar_c} by considering the following linear
constraint
\begin{align}\label{eq:Linrz}
\tr(\Wb_k\Qb_{ki})\le{e}^{\bar{x}_{ki}}(x_{ki}-\bar{x}_{ki}+1),
\end{align} for $k\in \mathcal{K}^c_{i}$. To approximate
\eqref{WSR_ChVar_d}, we consider the following lower bound for
$e^{y_i}+1$:
\begin{align}\label{monomial}
\left(\frac{e^{y_i}}{{\theta}_{i1}(\bar{y}_i)}\right)^{{\theta}_{i1}(\bar{y}_i)}\left(\frac{1}{{\theta}_{i2}(\bar{y}_i)}\right)^{{\theta}_{i2}(\bar{y}_i)}\le{e}^{y_i}+1,
\end{align}
where ${\theta}_{i1}(\bar{y}_i)={e^{\bar{y}_i}}/({e^{\bar{y}_i}+1})$
and ${\theta}_{i2}(\bar{y}_i)={1}/({e^{\bar{y}_i}+1})$. Equation
\eqref{monomial} is obtained from the inequality of arithmetic and
geometric means. By \eqref{monomial}, a sufficient condition for
\eqref{WSR_ChVar_d} can be obtained as
\begin{align}\label{eq:condensation}
(\theta_{i1}(\bar{y}_i))^{\theta_{i1}(\bar{y}_i)}(\theta_{i2}(\bar{y}_i))^{\theta_{i2}(\bar{y}_i)}e^{(\ln{2})R_i-\theta_{i1}(\bar{y}_i)y_i}\le1,
\end{align}
for $i=1,\dots,K$, which are convex constraints. By replacing
\eqref{WSR_ChVar_c} and \eqref{WSR_ChVar_d} with \eqref{eq:Linrz}
and \eqref{eq:condensation}, respectively, and by ignoring the
nonconvex rank-one constraints in \eqref{WSR_ChVar_g}, we obtain the
following conservative formulation for problem \eqref{WSR_ChVar}:
\begin{center}
\fbox{\parbox[]{0.97 \linewidth}{\vspace{-0.3cm}
\begin{align}\label{WSR_ChVar2}
\max_{\substack{\Wb_i \in \mathbb{H}^{N_t},R_i\geq 0, \\ x_{ki},y_i,z_i \in \mathbb{R}, \\k,i=1,\ldots,K}}~&\sum_{i=1}^K\alpha_iR_i, \\
\text{s.t.}~&~\rho_ie^{\sigma_i^2z_i}\prod_{k\ne{i}}\left(1+e^{-x_{ii}+x_{ki}+y_i}\right)\leq 1,\notag\\
&~\tr(\Wb_k\Qb_{ki})\le{e}^{\bar{x}_{ki}}(x_{ki}-\bar{x}_{ki}+1),~~k\in \mathcal{K}^c_i,\notag\\
&~\tr(\Wb_i\Qb_{ii})\geq e^{x_{ii}},\notag\\
&~\Theta_i(\bar{y}_i)e^{(\ln{2})R_i-\theta_{i1}(\bar{y}_i)y_i}\le1,\notag\\
&~e^{y_i-x_{ii}}\leq z_i,\notag\\
&~\tr(\Wb_i)\leq P_i,~\Wb_i\succeq \zerob,~i=1,\dots,K, \notag
\end{align}
\vspace{-0.6cm}}}
\end{center}
%\end{subequations}
where $\Theta_i(\bar{y}_i) \triangleq
(\theta_{i1}(\bar{y}_i))^{\theta_{i1}(\bar{y}_i)}(\theta_{i2}(\bar{y}_i))^{\theta_{i2}(\bar{y}_i)}$.
Problem \eqref{WSR_ChVar2} is a convex optimization problem; it can
be efficiently solved by standard convex solvers such as
\texttt{CVX} \cite{cvx}.

The idea of removing the nonconvex rank-one constraints of
$\{\Wb_i\}_{i=1}^K$ in \eqref{WSR_ChVar2} is known as semidefinite
relaxation (SDR) in convex optimization theory \cite{Luo2010_SPM}.
SDR is in general an approximation because the optimal
$\{\Wb_i\}_{i=1}^K$ of problem \eqref{WSR_ChVar2} may not be of rank
one. Surprisingly, it is found in our simulations that problem
\eqref{WSR_ChVar2} always yields rank-one optimal
$\{\Wb_i\}_{i=1}^K$, i.e., $\Wb_i=\wb_i(\wb_i)^H$ for all $i$,
provided that $\Wb_i \neq \zerob$. This implies that, for the
problem instances we tested in simulations, an approximate
beamforming solution to problem \eqref{WSR} can be directly obtained
by decomposing the optimal $\{\Wb_i\}_{i=1}^K$ of problem
\eqref{WSR_ChVar2}.

\vspace{-0.3cm}
%===========================================================================================
\subsection{Sequential Conservative Approximations}\label{sec:IteApproach}
%===========================================================================================
\vspace{-0.1cm}

The formulation \eqref{WSR_ChVar2} is obtained by conservatively
approximating problem \eqref{WSR} with respect to the feasible point
$(\{\bar{\wb}_i\}_{i=1}^K,\{\bar{R}_i\}_{i=1}^K)$ [see
\eqref{feasible point}]. It is possible to further improve the
approximation performance by solving problem \eqref{WSR_ChVar2}
iteratively with the optimal
$(\{{\wb}_i\}_{i=1}^K,\{{R}_i\}_{i=1}^K)$ at the current iteration
used as the feasible point
$(\{\bar{\wb}_i\}_{i=1}^K,\{\bar{R}_i\}_{i=1}^K)$ for the next
iteration. The proposed sequential approximation algorithm is
summarized in the following Algorithm 1: \vspace{-0.2cm}
\begin{algorithm}[h]
  \caption{Proposed sequential conservative approximation algorithm for solving problem \eqref{WSR}}
\begin{algorithmic}[1]
  \State {\bf Input} a feasible point
  $(\{\bar{\wb}_i\}_{i=1}^K,\{\bar{R}_i\}_{i=1}^K)$ of problem
  \eqref{WSR}, and a solution accuracy $\delta>0$.
  \State {Obtain} $\{\{\bar{x}_{ki}\}_{k\neq i},\bar{y}_i\}_{i=1}^K$ by \eqref{feasible point} and obtain
        ${\theta}_{i1}(\bar{y}_i)={e^{\bar{y}_i}}/({e^{\bar{y}_i}+1})$
        and ${\theta}_{i2}(\bar{y}_i)={1}/({e^{\bar{y}_i}+1})$ for
        $i=1,\ldots,K.$
  \State {Solve} problem \eqref{WSR_ChVar2} to obtain the
  optimal beamforming matrices $\{\Wb_i^\star\}_{i=1}^K$ and rates
  $\{R_i^\star\}_{i=1}^K$.

  \State Obtain $\wb_i^\star$ by decomposition of
  $\Wb_i^\star=\wb_i^\star(\wb_i^\star)^H$ for $i=1,\ldots,K$.

  \State {\bf Output} the approximate beamforming solution
  $(\wb_1^\star,\ldots,\wb_K^\star)$ and achievable rate tuple $(R_1^\star,\ldots,R_K^\star)$
  if $|\sum_{i=1}^K\alpha_i R_i^\star-\sum_{i=1}^K\alpha_i \bar{R}_i|/\sum_{i=1}^K\alpha_i
  \bar{R}_i<\delta$; otherwise update $\bar{\wb}_i:=\wb_i^\star$ and
  $\bar{R}_i:=R_i^\star$ for all $i$, and go to Step 2.
\end{algorithmic}
\end{algorithm}
\vspace{-0.2cm}

A feasible point to initialize Algorithm 1 can be easily obtained
by some heuristic transmission strategies. For example, one can
obtain a feasible point
$(\{\bar{\wb}_i\}_{i=1}^K,\{\bar{R}_i\}_{i=1}^K)$ of problem
\eqref{WSR} through the simple maximum-ratio transmission (MRT)
strategy. In this strategy, the beamforming vectors
$\{\bar{\wb}_i\}_{i=1}^K$ are simply set to $\bar{\wb}_i =
\sqrt{P_i} \qb_i$ where $\qb_i \in \Cplx^{N_t}$, $\|\qb_i\|=1$, is
the principal eigenvector of $\Qb_{ii}$ for $i=1,\ldots,K$. For
the $i$th transmitter-receiver pair, the associated
$\epsilon_i$-outage achievable rate of MRT is given by the maximum
$\bar{R}_i$ that satisfies the following inequality [see
\eqref{WSR_CLSFORM}]
\begin{align*}
\rho_i
e^{\frac{(2^{\bar{R}_i}-1)\sigma_i^2}{\bar{\wb}_i^H\Qb_{ii}\bar{\wb}_i}}\prod_{k\ne{i}}
\left(\!1\!+\!\!\frac{(2^{\bar{R}_i}-1)\bar{\wb}_k^H\Qb_{ki}\bar{\wb}_k}{\bar{\wb}_i^H\Qb_{ii}\bar{\wb}_i}\right)\le1.
\end{align*}
Analogously, one can also obtain a feasible point of \eqref{WSR} by
the zero-forcing (ZF) transmission strategy, provided that $N_t$ is
sufficiently large. In the next section, we present some simulation
results to demonstrate the efficacy of the proposed approximation
algorithm.

\vspace{-0.1cm}
%===========================================================================================
\section{Simulation Results and Discussions}\label{sec:SimuRslt}
%===========================================================================================
\vspace{-0.2cm}

In the simulations, we consider the multiuser MISO-IFC as described
in Section 2. For simplicity, all the receivers are assumed to have
the same noise power, i.e., $\sigma_1^2=\cdots=\sigma_K^2 \triangleq
\sigma^2$, and all the power constraints are set to one, i.e.,
$P_1=\cdots=P_K=1$. The channel covariance matrices $\Qb_{ki}$ were
randomly generated. We normalize the maximum eigenvalue of
$\Qb_{ii}$, i.e., $\lambda_{\max}(\Qb_{ii})$, to one for all $i$,
and normalize $\lambda_{\max}(\Qb_{ki})$ to a value $\eta  \in(0,1]$
for all $k\in
 \mathcal{K}^c_i$, $i=1,\ldots,K$. The parameter $\eta$, thereby, represents the relative cross-link interference
level. If not mentioned specifically, the ranks of $\Qb_{ki}$ are
all set to $N_t$. We consider the sum rate maximization problem by
setting $\alpha_1=\cdots=\alpha_K=1$ for problem \eqref{WSR}, and
set $\epsilon_1=\cdots=\epsilon_K=0.1$, i.e., $10\%$ outage
probability. For the proposed approximation algorithm (Algorithm 1),
we set $\delta=10^{-2}$ and use \texttt{CVX} \cite{cvx} to handle
the associated problem \eqref{WSR_ChVar2}. All the simulation
results were obtained by averaging over $500$ trials.

In the first example, we examine the approximation accuracy of the
proposed method by comparing with the optimal sum rate obtained by
the exhaustive search method in \cite{Lindblom09}. Figure 1 shows
the simulation results of average achievable rate versus $\eta$ for
$K=2$ and $N_t=4$. The achievable rate of the simple TDMA scheme is
also shown in this figure. Firstly, one can see from this figure
that the sum rate achieved by the proposed method approaches that of
TDMA with increased $\eta$; TDMA exhibits a constant sum rate for
all $\eta$ because there is no cross-link interference for this
scheme. Secondly, we observe that the proposed method can exactly
attain the average optimal sum rate for $1/\sigma^2=0$ dB and
$1/\sigma^2=10$ dB. For $1/\sigma^2=20$ dB and for $\eta\geq 0.5$
(interference dominated scenarios), it can be observed that there is
a small gap between the rate achieved by the proposed method and the
optimal rate. Nevertheless, this gap is within $3\%$ of the optimal
sum rate on average.

In the second example, we compare the proposed method with the MRT
scheme and TDMA for $N_t=K=4$. Figure 2 shows the results of
average sum rate versus $1/\sigma^2$. Note that, for the case of
$K=4$, the exhaustive search method in \cite{Lindblom09} is too
complex to implement, and thus no result for the optimal sum rate
is shown. From Fig. 2, we can observe that the proposed method
achieves the highest sum rate among the three methods, no matter
when $\eta=0.2$ or $\eta=1$. One can also see that, for $\eta=0.2$
and $1/\sigma^2< 5$ dB, MRT can yield a sum rate comparable to the
proposed method and outperforms TDMA; whereas TDMA performs better
for $\eta=1$.

In order to compare with the ZF scheme, in the third example, we
extend the number of antennas to 8 ($N_t=8$) and constrain the
ranks of all channel covariance matrices to 2. The simulation
results are shown in Fig. 3. As seen from this figure, the
proposed method still performs best compared to the other three
schemes. On the other hand, one can see that ZF can achieve a
higher average sum rate than TDMA, and also outperforms MRT for
$\eta=1$.

\begin{figure}[t]
\begin{center}
\includegraphics[scale=0.45]{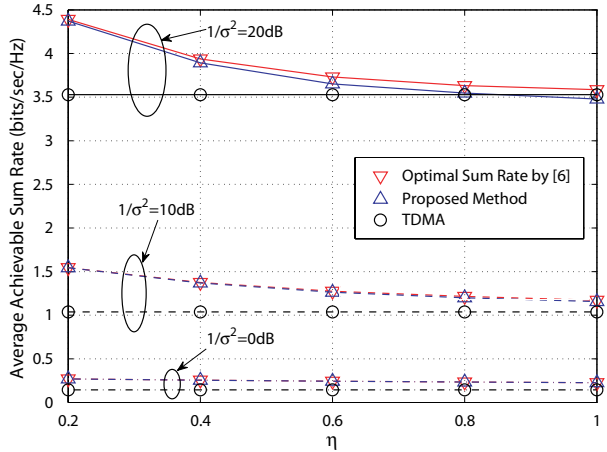}
\end{center}\vspace{-0.8cm}
\caption{Average achievable sum rate versus $\eta$ for $K=2$,
$N_t=4$, and $\rank(\Qb_{ki})=4$ for all
$k,i$.}\label{fig:fig1}\vspace{-0.4cm}
\end{figure}
%%%%%%%%%%%%%%%%%%%%%%%%%%%%%%%%%%%%%%%%%%%%%%%%%%%%%%%%%%%%%%%%%%%%%%%%%%%%%%%%%%%%%%%%%%%%

%%%%%%%%%%%%%%%%%%%%%%%%%%%%%%%%%%%%%%%%%%%%%%%%%%%%%%%%%%%%%%%%%%%%%%%%%%%%%%%%%%%%%%%%%%%%
\begin{figure}[t]
\begin{center}
\includegraphics[scale=0.45]{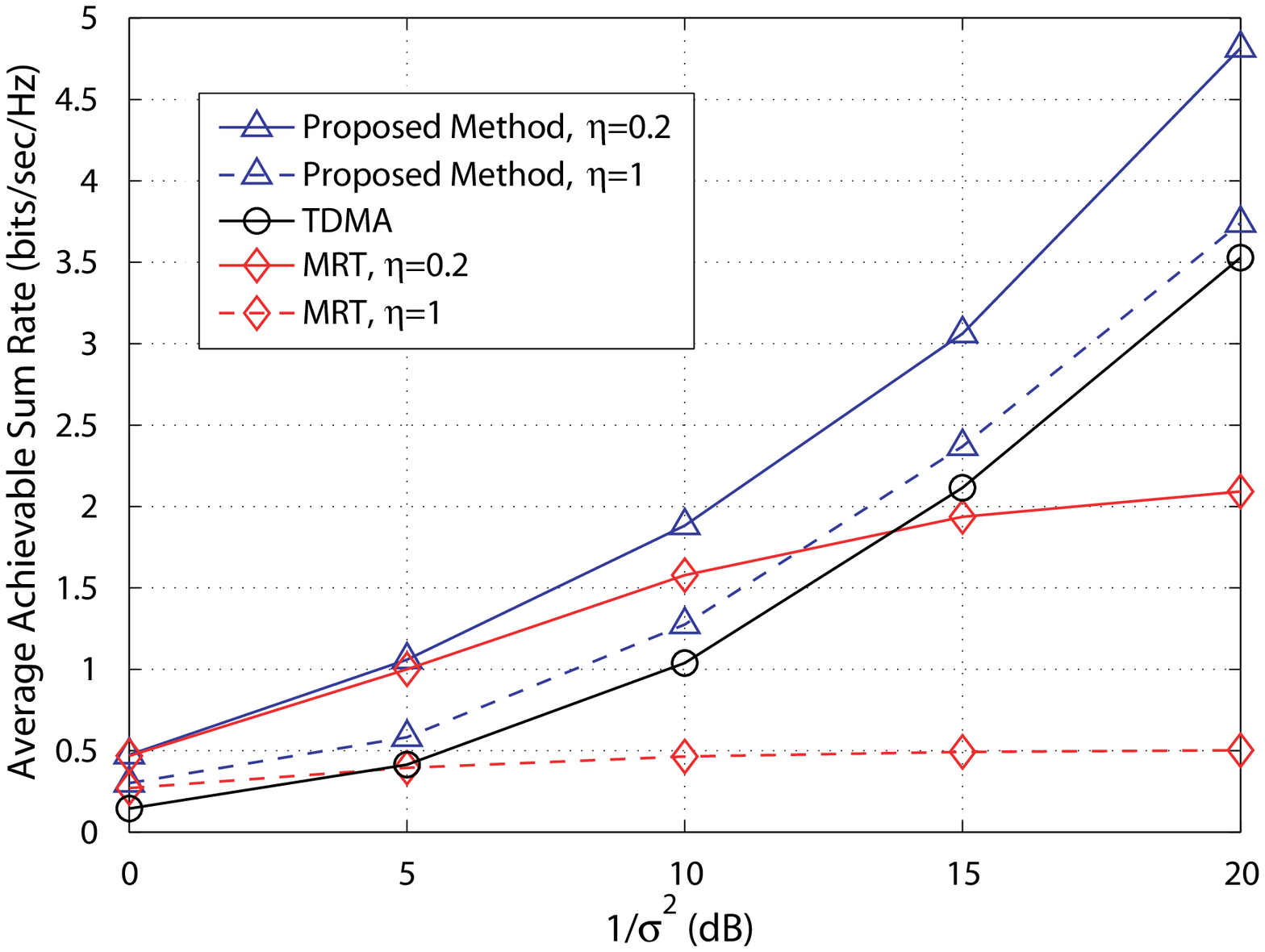}
\end{center}\vspace{-0.8cm}
\caption{Average achievable sum rate versus $1/\sigma^2$ for
$K=N_t=4$, and $\rank(\Qb_{ki})=4$ for all
$k,i$.}\label{fig:fig2}\vspace{-0.4cm}
\end{figure}
%%%%%%%%%%%%%%%%%%%%%%%%%%%%%%%%%%%%%%%%%%%%%%%%%%%%%%%%%%%%%%%%%%%%%%%%%%%%%%%%%%%%%%%%%%%%

%%%%%%%%%%%%%%%%%%%%%%%%%%%%%%%%%%%%%%%%%%%%%%%%%%%%%%%%%%%%%%%%%%%%%%%%%%%%%%%%%%%%%%%%%%%%
\begin{figure}[t]
\begin{center}
{\includegraphics[scale=0.45]{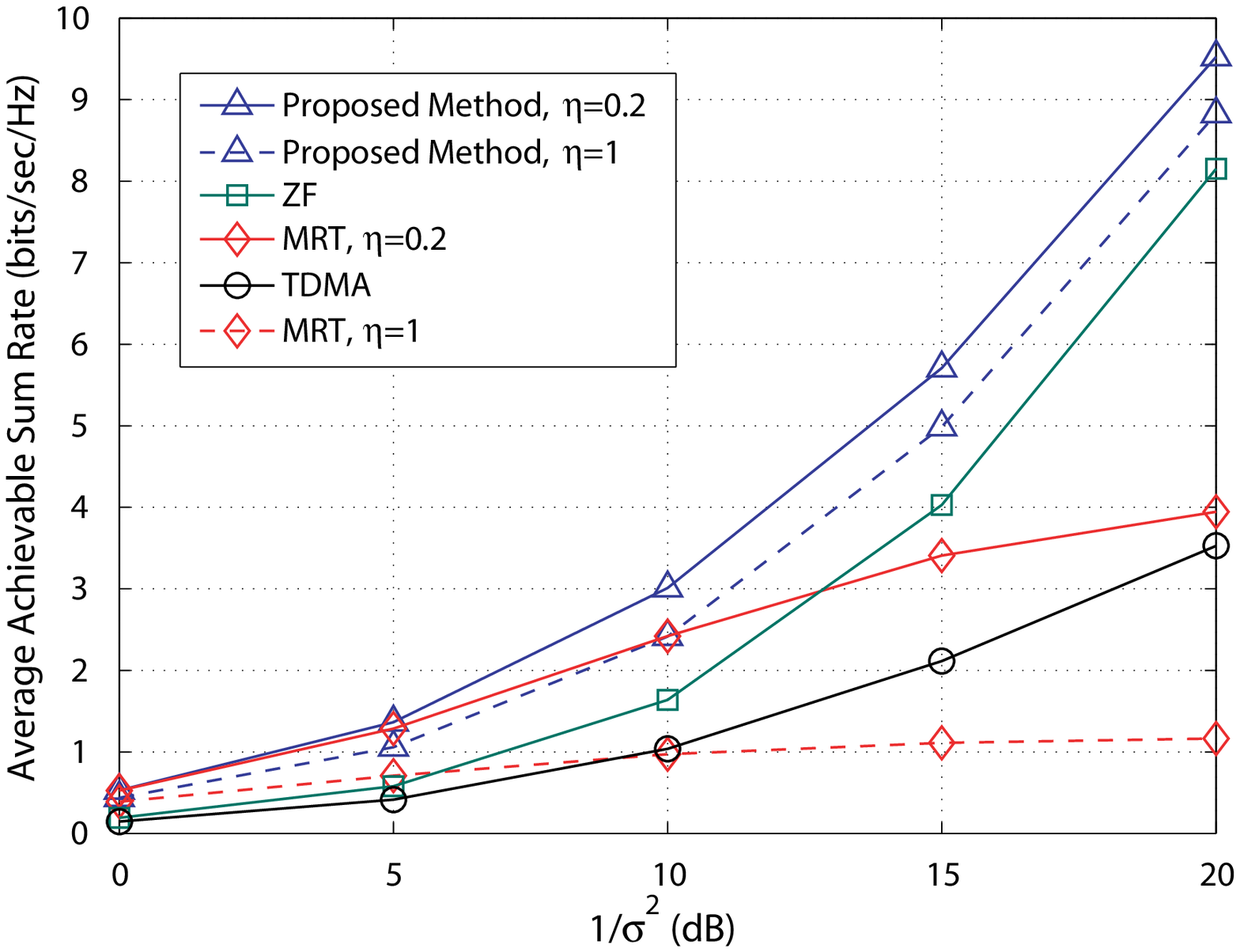}}
\end{center}\vspace{-0.8cm}
\caption{Average achievable sum rate versus $1/\sigma^2$ for $K=4$,
$N_t=8$, and $\rank(\Qb_{ki})=2$ for all
$k,i$.}\label{fig:fig3}\vspace{-0.4cm}
\end{figure}
%%%%%%%%%%%%%%%%%%%%%%%%%%%%%%%%%%%%%%%%%%%%%%%%%%%%%%%%%%%%%%%%%%%%%%%%%%%%%%%%%%%%%%%%%%%%
%%===========================================================================================

%===========================================================================================
\vspace{-0.2cm} \footnotesize
\bibliography{IFC_references}

\begin{thebibliography}{1}
\providecommand{\url}[1]{#1}
\csname url@samestyle\endcsname
\providecommand{\newblock}{\relax}
\providecommand{\bibinfo}[2]{#2}
\providecommand{\BIBentrySTDinterwordspacing}{\spaceskip=0pt\relax}
\providecommand{\BIBentryALTinterwordstretchfactor}{4}
\providecommand{\BIBentryALTinterwordspacing}{\spaceskip=\fontdimen2\font plus
\BIBentryALTinterwordstretchfactor\fontdimen3\font minus
  \fontdimen4\font\relax}
\providecommand{\BIBforeignlanguage}[2]{{%
\expandafter\ifx\csname l@#1\endcsname\relax
\typeout{** WARNING: IEEEtran.bst: No hyphenation pattern has been}%
\typeout{** loaded for the language `#1'. Using the pattern for}%
\typeout{** the default language instead.}%
\else
\language=\csname l@#1\endcsname
\fi
#2}}
\providecommand{\BIBdecl}{\relax}
\BIBdecl

\bibitem{Cadambe08}
V.~R. Cadambe and S.~A. Jafar, ``Interference alignment and degrees of freedom
  of the d-user interference channel,'' \emph{IEEE Trans. Inf. Theory},
  vol.~54, pp. 3425--3441, Aug. 2008.

\bibitem{Shang_2009}
X.~Shang, B.~Chen, and H.~V. Poor, ``On the optimality of beamforming for
  multi-user {MISO} interference channels with single-user detection,'' in
  \emph{Proc. IEEE GLOBECOM}, Honolulu, Hawaii, USA, Nov. 30-Dec. 4, 2009, pp.
  1--5.

\bibitem{Jorswieck08}
E.~A. Jorswieck, E.~G. Larsson, and D.~Danev, ``Complete characterization of
  the {Pareto} boundary for the {MISO} interference channel,'' \emph{IEEE
  Trans. Signal Process.}, vol.~56, pp. 5292--5296, July 2008.

\bibitem{Zhang_Cui_2010}
R.~Zhang and S.~Cui, ``Cooperative interference management with {MISO}
  beamforming,'' \emph{IEEE Trans. Signal Process.}, vol.~58, pp. 5450--5458,
  Oct. 2010.

\bibitem{Larsson_etal2009_mag}
E.~G. Larsson, E.~A. Jorswieck, J.~Lindblom, and R.~Mochaourab, ``Game theory
  and the flat-fading {Gaussian} interference channel,'' \emph{IEEE Signal
  Process. Mag.}, pp. 18--27, Sept. 2009.

\bibitem{Lindblom09}
J.~Lindblom, E.~Karipidis, and E.~G. Larsson, ``Outage rate regions for the
  {MISO IFC},'' in \emph{Proc. 43rd Asilomar Conference}, Pacific Grove, CA,
  Nov. 1-4, 2009, pp. 1120--1124.

\bibitem{cvx}
M.~Grant and S.~Boyd, ``{CVX}: {M}atlab software for disciplined convex
  programming,'' http://stanford.edu/$\sim$boyd/cvx, June 2009.

\bibitem{Kandukuri02}
S.~Kandukuri and S.~Boyd, ``Optimal power control in interference-limited
  fading wireless channels with outage-probability specifications,'' \emph{IEEE
  Trans. Wireless Commun.}, vol.~1, pp. 46--55, Jan. 2002.

\bibitem{Luo2010_SPM}
Z.-Q. Luo, W.-K. Ma, A.~M.-C. So, Y.~Ye, and S.~Zhang, ``Semidefinite
  relaxation of quadratic optimization problems,'' \emph{IEEE Signal Process.
  Mag.}, pp. 20--34, May 2010.

\end{thebibliography}
%===========================================================================================

\end{document}